\documentstyle[twocolumn,aps,prl,epsf,floats]{revtex}
\begin{document} 
\twocolumn[\hsize\textwidth\columnwidth\hsize\csname @twocolumnfalse\endcsname
\title {
Temperature variation of the pseudogap in
 underdoped cuprates}
\author{Andrey V. Chubukov$^1$ and J\"org Schmalian$^2$}
\address{
$^1$ Department of Physics, University of Wisconsin, Madison, WI 53706\\
$^2$ Department of Physics, University of Illinois at Urbana-Champaign, Urbana, IL 61801}
\date{October 20, 1997}
\maketitle
\begin{abstract}
We consider thermal evolution of the 
 spectral function $A(\Omega, T)$ in underdoped cuprates at 
$T >T_{\rm c}$. We find that 
in the strong coupling limit, the
fermionic Green's function near $(0,\pi)$ behaves as 
$G^{-1} (\Omega) \propto \sqrt{\Omega}$ in the whole
 frequency range relevant to
photoemission experiments. The analysis of the pairing problem with this form
of $G$ yields
a leading edge gap in $A(\Omega)$ and a broad maximum at larger
frequencies. We find that classical fluctuations predominantly add
a frequency independent term to $G^{-1} (\Omega)$. This
term destroys the leading edge gap at $T =T^* ({\bf k})$ which
has a maximum value of about $200K$                
near $(0,\pi)$. 
\end{abstract}
]
The pseudogap phenomenon in the normal state properties of 
underdoped cuprates is one of the key issues
 in the field of high-$T_{\rm c}$ superconductivity. 
NMR, transport, optical and photoemission measurements~\cite{review}
 all demonstrate that underdoped cuprates behave in many respects like a 
$d-$wave  superconductor at temperatures which can be several times 
larger than actual $T_{\rm c}$.
In photoemission measurements, the evidence for a pseudogap
 is based on the
observation of  a leading edge gap (LEG) 
in the quasiparticle
spectral function $A(\Omega)$ near $(0,\pi)$ and symmetry related
points~\cite{pseudo,Norman}. 
Contrary to situation below $T_{\rm c}$, however, the LEG  is not 
accompanied by a  sharp quasiparticle peak above it.
Instead, the spectral
function is flat above the LEG and only
displays a broad maximum at much larger frequencies~\cite{pseudo}. 
This form of $A(\Omega)$ exists
between $T_{\rm c}$, which can be very small in strongly 
underdoped cuprates, and some higher temperature, $T^* (k)$ which is 
momentum dependent~\cite{Norman} and 
has a maximum value of about $200K$ near $(0,\pi)$.
Similar observations have been made recently in 
tunneling experiments~\cite{tunnel}.

In a recent paper~\cite{AC}, one of us argued that 
 the LEG naturally appears in the 
in the ``magnetic" scenario for cuprates
 due to strong quantum magnetic 
fluctuations which give rise 
 to $d-$wave precursors  above $T_{\rm c}$, but  simultaneously almost
destroy the quasiparticle peak near $(0,\pi)$.
 These two effects yield a
LEG without a quasiparticle peak, in  agreement with the data. 

In this paper we show that, as temperature increases, 
magnetic fluctuations  loose their dynamics and 
progressively acquire a  classical, quasistatic behavior. We show that
this effect  acts against 
$d-$wave precursors. We further show that 
the  LEG is filled in by excitations
at temperatures which are comparable to a typical
spin-fluctuation frequency 
 and also depend on the momentum, being largest near $(0,\pi)$.

As in earlier studies~\cite{AC,Morr,joerg}, our consideration is based on the 
spin-fermion model
in which itinerant fermions which form a Fermi liquid
with a large, Luttinger-type Fermi surface 
interact with their collective spin 
degrees of freedom  by 
\begin{equation}
{\cal H}_{\rm s-f} = g 
\sum_{{\bf k}{\bf q}, \gamma \delta} c^{\dagger}_{{\bf k},\gamma} 
{\vec \sigma}_{\gamma,\delta}
c_{{\bf k}+{\bf q},\delta} {\vec S}_{-{\bf q}}\, .
\label{1}
\end{equation}
Here $g$ is the coupling constant  which is assumed to increase
as the system approaches half-filling, 
$\sigma_i$ are the Pauli matrices, and ${\vec S}_{\bf q}$ are collective spin variables
which  are characterized by their full spin susceptibility
$\chi_{\gamma \delta} (q, \omega) = \delta_{\gamma \delta} 
{\tilde \chi}/(1 + ({\tilde {\bf  q}} \xi)^2  -
i \omega/\omega_{\rm sf} - \omega^2/\Delta^2)$. Here $\xi$ is the
correlation length, $\Delta \propto \xi^{-1}$, and 
 ${\tilde {\bf q}} = {\bf Q} -{\bf q}$ where ${\bf Q}$ is the momentum at which the static
susceptibility is peaked. At small doping, ${\bf Q}$ is 
either equal or very close to the
antiferromagnetic momentum $(\pi,\pi)$. Further,
${\tilde \chi} = a \xi^2$ is the static susceptibility at ${\bf q}={\bf Q}$ and
$\omega_{\rm sf}$ is the typical frequency of overdamped spin fluctuations.
A fit to NMR experiments yields $a \sim 25$ states/eV,
$\omega_{\rm sf} \sim 10\, {\rm  meV}$ and $\xi \sim 3-4$ at the onset temperature
for a LEG~\cite{barz}. The value of $\Delta$ is unknown, 
but since $\omega^2$ term in $\chi$ accounts for
spin waves, it is reasonable to estimate $\Delta \sim v_s \xi^{-1}$
where $v_s$ is the spin-wave velocity.

Our goal is to obtain the form of the quasiparticle spectral function
$A({\bf k}, \Omega) = (1/\pi) |{\rm Im}~ G ({\bf k}, \Omega)|$ where $G^{-1} ({\bf k}, \Omega) =
G^{-1}_0 ({\bf k}, \Omega) - \Sigma ({\bf k}, \Omega)$. 
This can be most straightforwardly done by first evaluating
$\Sigma ({\bf k}, \Omega_m)$ for discrete Matsubara frequencies
 $\Omega_m = \pi T(2 m +1)$,
 and then performing analytical continuation to real frequencies.
To avoid  unnecessary complications, we also
restrict our consideration to the 
points in momentum space ${\bf k} = {\bf k}_{\rm hs}$ where
$\epsilon_{\bf k} = \epsilon_{{\bf k}+{\bf Q}} = \mu$, and the self-energy has the largest 
value. In cuprates
these points (hot spots) are located near $(0,\pi)$ and symmetry related
points. Finally, we restrict our consideration to second-order term in $g$ -
previous studies  indicate that
the second-order result for the fermionic self-energy  is dominant even
for large values of the coupling constant~\cite{Morr}. 

To second order in $g$, $\Sigma ({\bf k}, \Omega_m)$ is given by
$$\Sigma ({\bf k}, \Omega_m) = 3 g^2T
 \sum_{{\bf q}n}\,  
G_0 ({\bf k}+{\bf q},\Omega_m + \omega_n)~\chi_{\gamma \gamma} ({\bf q},\omega_n)$$
where $\omega_n = 2 \pi n T$ and $G_0$ is a bare fermionic Green's function
which near the Fermi surface has a form
$G_0 ({\bf k},\omega_n) = Z_0 /(i\omega_n -
\epsilon_{\bf k})$. Expanding in $\epsilon_{{\bf k}+{\bf q}}$
 to linear order in a
deviation from a hot spot and performing integration over momentum we obtain
\begin{eqnarray}
&&\Sigma ({\bf k}_{\rm hs}, \Omega_m) = - 2 i T \left(\frac{g}{g_0}\right)^2
 \times \nonumber \\
&&\sum_n 
\frac{\Omega_m + \omega_n}{|\Omega_m + \omega_n|}~K_n~{\rm cot}^{-1} 
\left(\frac{K_n|\Omega_m + \omega_n|}{v \xi^{-1}}\right)
\label{fullse}
\end{eqnarray}
where $K^{-2}_n = 1 + |\omega_n|/\omega_{\rm sf} + \omega^2_n/\Delta^2
-(( \Omega_m + \omega_n)/(v \xi^{-1}))^2$, and 
we introduced $g^2_0 = 4\pi v \xi^{-1}/(3 a Z^2_0)$ where $v$ is the
Fermi velocity at a hot spot.
The fits to photoemission and NMR data above $T^*$ yield, respectively,
 $v \sim 0.4 eV$ and  $v \xi^{-1}/\omega_{\rm sf} \sim 10$ 
independent on temperature.
For definiteness, we will also choose $v \xi^{-1}/\Delta =1$. 
 We have done calculations for larger 
ratios ($2$ and $4$) and obtain little difference in the results.

The summation over $n$ can be performed explicitly for $\Omega_m \ll v
\xi^{-1}$. For these $\Omega_m$, the use of the
 Poisson
summation formula yields
$$ \Sigma ({\bf k}_{\rm hs}, \Omega_m) = - i \left(\frac{g}{g_0}\right)^2 
\left( \frac{2  \Omega_m}{1 + \sqrt{1
 +\frac{|\Omega_m|}{\omega_{\rm sf}}}} + \Lambda (\Omega_m, T) \right)
$$ where $\Lambda(\Omega_m, T)$ is expressed in terms of Fresnel integrals.
The first term in this formula is
 a quantum result while the second term 
accounts for thermal
effects. At low $T \leq \omega_{\rm sf}$, the use of the asymptotic form of
Fresnel integrals yields
\begin{equation}
\Lambda(\Omega_m, T) = {\rm sign} \Omega_m~ \frac{(\pi T)^2}{4\omega_{\rm sf}} 
f\left(\frac{\Omega_m}{\omega_{\rm sf}}\right)\, ,
\label{lambda}
\end{equation}
where $f(0) = 1 -0.32 (\pi T/\omega_{\rm sf})^2 +....$ and $f(x\gg 1) =
(2/3)(1 - 0.28 (\pi T/\omega_{\rm sf})^2 + ...$ . 
The leading term in the expansion obviously 
yields  a conventional Fermi-liquid result for the self-energy. 
We see, however, that the typical scale
above which Fermi-liquid expansion breaks down 
 for both
external frequency and temperature is given by 
$\omega_{\rm sf}$ which is rather small in cuprates.
 In the
opposite limit of ``high" $T,\Omega \gg \omega_{\rm sf}$ (but still, 
$T,\Omega \ll v \xi^{-1}$),
 the self-energy has the form
\begin{equation}
 \Sigma ({\bf k}_{\rm hs}, \Omega) = - 2 \left(\frac{g}{g_0}\right)^2 
(\Omega \omega_{\rm sf})^{1/2} (1 + i \alpha(\Omega))\, ,
\label{oldse}
\end{equation}
where $\alpha(\Omega) = 1 + \beta \sqrt{\omega_{\rm sf}/\Omega}$,
 and $\beta = \pi T/(2\omega_{\rm sf})$. In this limit, the spin 
fluctuations are essentially static,
the thermal part of $ \Sigma ({\bf k}_{\rm hs}, \Omega)$
dominates, and the self-energy is linear in $T$. We however studied
the crossover between quadratic and linear in $T$ behavior of $\beta$ 
in more detail and found that
although  the deviations from the Fermi-liquid behavior
 begin at rather small $T$,
the asymptotic, linear in $T$ form of $\beta$
 is recovered only at $T \gg \omega_{\rm sf}$, 
i.e, the crossover
region is numerically rather wide. 

We will show below
that the onset temperature for the LEG falls into the intermediate
temperature range where quantum and classical contributions to the self-energy
are of about the same strength. But before, we briefly review the earlier
results~\cite{AC} of a purely quantum consideration when $\beta =0$
which was argued to give rise to a LEG without a quasiparticle peak.
The key points of the quantum consideration are:

(i) The 
self-energy as in (\ref{oldse}) survives in 
some region around a hot spot with the
width in ${\bf k}$-space $\sim \Omega/\omega_{\rm sf} \xi^2$.
Numerical estimates show that for $\Omega \geq \omega_{\rm sf}$, this region
is rather wide and, e.g., for ${\bf Q}=(\pi,\pi)$ it includes
$(0,\pi)$ and symmetry related points.

(ii)When $g$ becomes few times larger than $g_0$ 
(which is likely to happen when the system
approaches optimal doping from the overdoped side),
the self-energy overshadows the bare fermionic dispersion such that $G^{-1}
({\bf k}, \Omega) \approx \Sigma ({\bf k}_{\rm hs}, \Omega)$. 
As real and imaginary parts of $\Sigma ({\bf k}_{\rm hs}, \Omega)$ in (\ref{oldse}) are
the same for $\beta =0$, the dominance of the self-energy over bare dispersion
leads to a transform of a spectral weight to higher frequencies,
and to a near-destruction of the 
quasiparticle peak near hot spots.

\begin{figure} [t]
\begin{center}
\leavevmode
\epsffile{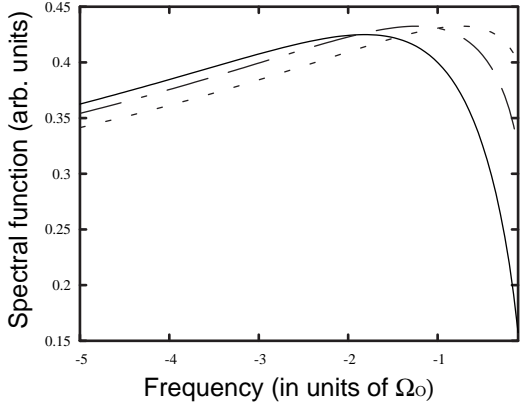}
\end{center}
\caption{The plot of the spectral function $A(\Omega)$ given by Eq.
(\protect\ref{sf}). Solid, dashed-dotted and dotted lines correspond to
${\protect\bar \beta}  
=0,~ 1/2$ and $1$, respectively. The spectral function for 
${\protect\bar \beta} =0$ displays
the LEG and the broad maximum at higher frequencies. As ${\protect\bar \beta}$
increases with temperature, the LEG is progressively filled in by excitations.}
\label{fig1}
\end{figure}

(iii) The analysis of the pairing susceptibility shows
that though the $d-$wave
pairing susceptibility  turns out to be substantially enhanced for $g > g_0$,
the near-destruction of the quasiparticle peak  prevents
the system to become an actual superconductor down to very small $T$.
At the same time,  due to the enhancement and narrowing of the pairing susceptibility
at $g > g_0$, the frequency shift which one  obtains
by solving the Gorkov equation for the
full Green's function in the pairing channel becomes larger than
 the width of the pairing susceptibility. 
In this situation, the solution of the Gorkov's equation has almost the same
form as in the actual superconducting state where the pairing susceptibility
has a $\delta-$functional form, i.e., system develops precursors
to the $d-$wave superconducting state without developing a true 
superconductivity.
It is essential, however, that as $G^{-1} ({\bf k}, \Omega) \propto e^{i
\pi/4} \sqrt{\Omega}$,  the frequency shift due to  $d-$wave precursors 
yields a pole in the full $G$ along an imaginary rather than a
real frequency axis. This in turn yields the spectral function in the
form
\begin{equation}
A_{\beta =0} (\Omega) \propto \sqrt{\Omega}~
\frac{\Omega + \Omega_0}{\Omega^2 + \Omega^2_0}
\label{sfq}
\end{equation}
where explicit computations yielded~\cite{AC}
$\Omega_0 = \Omega_0 (k) \approx 0.68 \omega_{\rm sf} \xi^{2} (\cos k_x - \cos
k_y)^2$. 
This spectral function does not possess  a sharp quasiparticle pole, but
nevertheless it shows a ``gap-like" behavior. Indeed, $A(\omega)$
rapidly increases at low frequencies, reaches half
maximum at $\Omega \sim 0.2 \Omega_0$, then passes through a
broad maximum at $\Omega = \Omega_0$, and slowly decreases at
higher frequencies. This behavior, shown in Fig. ~\ref{fig1},
 is in a good agreement 
with the form of the photoemission curve observed in experiments below
$T^*$~\cite{pseudo}.

We now study how the spectral function changes as one takes thermal effects
into consideration. We assume that the conditions for $d-$wave
precursors are satisfied (i.e., the width of the pairing susceptibility is
smaller than the frequency shift obtained by solving the Gorkov's equation),
but the self-energy, and, hence inverse bare Green's function for the pairing
channel, has the form of Eq. (\ref{oldse}) with $\beta >0$. For the spectral
function, we then obtain instead of (\ref{sfq})
\begin{equation}
A (\Omega) \propto ~
\frac{\alpha (\Omega) \sqrt{\Omega} ~(\Omega\frac{1 + \alpha^2(\Omega)}{2} + 
\Omega_0)}{(\Omega\frac{1 + \alpha^2(\Omega)}{2})^2  + \Omega^2_0 - \Omega
\Omega_0 (1 - \alpha^2 (\Omega))}.
\label{sf}
\end{equation}
Here it is convenient to rewrite $\alpha (\Omega)$ as
$\alpha (\Omega) = 1 + {\bar \beta} \sqrt{\Omega_0/\Omega}$ where
${\bar \beta} = \beta/(0.82 \xi |cos k_x - cos k_y|)$. 
The plots of $A(\Omega)$ for various values of ${\bar \beta}$ 
are presented in
Fig.~\ref{fig1}. We see that the 
LEG is progressively filled in as $\bar \beta$ increases 
from $0.5$ to $1$. This happens chiefly
because thermal fluctuations eliminate the $\sqrt{\Omega}$ dependence
at small frequencies. Clearly, the temperature where the
LEG disappears depends on ${\bf k}$ being the largest near the points where the
$d-$wave gap has the largest value. 
The high-frequency part of $A(\Omega)$,
which possesses a broad maximum, is a bit more robust and 
disappears at somewhat larger ${\bar \beta}$.
 In other words, as temperature increases, thermal excitations
first fill the LEG,
and then, at somewhat higher $T$, destroy the broad maximum at
$\Omega_0$. This behavior seems to be consistent with both
photoemission~\cite{Norman} and
tunneling experiments~\cite{tunnel}.
Notice that this mechanism is {\it different} from a conventional
thermal smearing of the gap which in our approach would
would correspond to a thermal
broadening of the pairing susceptibility, $\chi^{sc} (\Omega)$.
\begin{figure} [t]
\begin{center}
\leavevmode
\epsffile{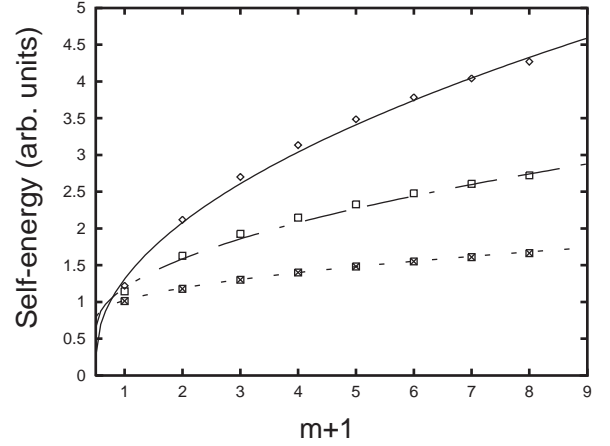}
\end{center}
\caption{The results for the full self-energy (\protect\ref{fullse})
for first $8$ Matsubara frequencies $\Omega_m = \pi T (2m+1)$ and the
fits to $\protect\sqrt{\Omega_m} +
 \beta \protect\sqrt{\omega_{sf}/2}$ dependence. 
Solid, dashed-dotted and dashed
lines correspond to $T =\omega_{\rm sf}/2,
 ~T=\omega_{\rm sf}$ and $T = 2 \omega_{\rm sf}$, respectively.}
\label{fig2}
\end{figure}
We now need to find the actual 
temperature dependence of $\beta$ and obtain the
onset temperature for the LEG which can  be directly compared with
the data. For this, we compute the self-energy numerically by performing 
the summation over $n$
 in (\ref{fullse}) without simplifying the form of the
integrand, i.e., without assuming that $\Omega_m \ll v \xi^{-1}$.
 We performed computations for $T = \omega_{\rm sf}/2,
 T = \omega_{\rm sf}$ and $T= 2 \omega_{\rm sf}$ and in all 
three cases found that the self-energy 
for external fermionic frequencies $\Omega_m = \pi T (2m +1)$ 
and $0\leq m \leq 7$ 
can be well approximated by the same form as in (\ref{oldse}), i.e.
$\Sigma ({\bf k}_{\rm hs}, 
 \Omega_m) \propto (\sqrt{\Omega_m} + \beta \sqrt{\omega_{\rm sf}/2}$.
The accuracy of the
fit is demonstrated in Fig.~\ref{fig2}. We emphasize that 
this procedure shows that the exact self-energy follows 
the $\sqrt{\Omega}$ dependence in a much larger frequency
range than one may expect from the  approximate analytical treatment.  
The overall factor in the fit 
decreases a bit with increasing temperature.
 This last renormalization, however,
can be absorbed into the
renormalization of $g$ and is irrelevant as
the positions of both the LEG and the broad maximum in $A(\Omega)$
are governed by $\Omega_0$, which does not depend on the coupling strength
as long as $g > g_0$.
 
We now present the numbers. We obtained $\beta =0.24 (0.38), 1.69 (2.57)$
 and $6.82 (10.63)$ for 
$v \xi^{-1}/\Delta =1 (2)$ and $T/\omega_{\rm sf} = 0.5, 1$ 
and $2$, respectively. 
We see that $\beta$ rapidly increases with increasing temperature.
The increase roughly follows $T^{\nu}$ dependence
 with $\nu$ somewhat larger than $2$,
i.e., it is much stronger than  the linear in $T$ increase
which one would expect for a purely classical self-energy. On the other
hand, already for $T =
\omega_{\rm sf}/2$, the value of $\beta$ is about $40\%$ smaller than in
the Fermi-liquid theory. This collaborates our earlier observation in
(\ref{lambda}) that the Fermi-liquid behavior is altered already at low $T$,
but there is a wide intermediate temperature range where the thermal part
of the self-energy interpolates between quadratic and linear in $T$ behavior.

Using  $\bar \beta =0.5$ as a criteria for the onset of the 
LEG, we find that the LEG near $(0,\pi)$ is 
filled in at $T \sim T^* =0.6 \omega_{\rm sf}
\xi^{1/\nu}$.
A fit to NMR shows that $T^* \sim 150 -200K$. Away from $(0,\pi)$,
the LEG is filled in at smaller temperatures which roughly follows
$T^* ({\bf k}) = T^* (|cos k_x - cos k_y|/2)^{1/\nu}$. 
Both of these results are in 
quantitative agreement agreement  
with the data~\cite{pseudo,Norman}.  In addition, as one moves
 towards the
 zone diagonal, i.e., goes away from the  hot spots,
 the overall strength of the $\sqrt{\Omega}$ term in the self-energy 
decreases. This pushes the system back towards
a weak coupling limit and causes  an additional decrease in $T^* ({\bf k})$.
Simultaneously, one should gradually recover 
some features of a conventional quasiparticle
peak in $A(\Omega)$.  This 
last effect has also been observed in recent experiments~\cite{Norman}. 
Finally, a fit to the NMR data
shows that  the ratio $T/\omega_{\rm sf}$ flattens at $\sim 1.4-1.5$ slightly
above $T^*$ and
remains constant upto much higher  $T$.
Then, $\bar \beta$ also 
flattens at a value $\sim 1$. This $\bar \beta$ still 
yields a broad maximum in $A(\Omega)$ 
above $T^*$, in agreement with the data.

We also emphasize that in our approach, the LEG 
gradually appears  at $T \sim T^* ({\bf k})$  without a 
break up 
of the Fermi surface. In other words, the present theory 
predicts that  
 above $T_{\rm c}$,
there still exists a quasiparticle peak at a 
vanishing frequency at ${\bf k} = {\bf k}_{\rm F}$, 
 The residue of the peak,
however, decreases as one moves away from the zone diagonal. 
Near $(\pi,0)$ it scales as $(g_0/g)^2$ which
seems too small to be detected by current  photoemission experiments.
Notice in this regard that photoemission data show that
the quasiparticle peak is clearly 
present near the zone diagonal, and {\it gradually}
disappears as one approaches $(0,\pi)$.
The  tunneling experiments~\cite{tunnel},
which  measure the momentum averaged 
spectral density,  also seem to observe a coherent peak above $T_c$
in the unoccupied part of the density of states.

A final note. In the present consideration, 
 the LEG and the broad maximum in $A(\omega)$ at a finite frequency are both
$d-$wave precursors in a Fermi liquid with strong magnetic fluctuations. 
A number of other scenarios
for a pseudogap have also been proposed~\cite{Morr,joerg,KS,lee}.
In particular, several groups 
argued~\cite{Morr,KS,joerg} that the broad
maximum in the spectral function  is a precursor of an antiferromagnetic SDW 
state. In the present calculations, SDW precursors do not
appear because  the fermionic self-energy is an
increasing function of frequency upto very large
frequencies. The SDW precursors, on the contrary,  can appear only
if the  self-energy decreases with frequency~\cite{Morr}. 
We found, however, that  the
form of the spectral function at
 $\Omega \geq v \xi^{-1}$ is highly sensitive 
to the high-frequency form of the 
spin susceptibility.
Namely,  
we did calculations
using  $\chi ({\bf q},\Omega)$ in which the $\omega^2$ term is substituted
by a hard cutoff at $\omega \sim \Delta$ and found that the 
classical behavior sets up at much smaller temperatures.   
In this last case, for $T > T^*$, the dominant contribution to
(\ref{fullse}) comes from the $n=0$ term, i.e., spin fluctuations are
essentially static. These fluctuations  yield the fermionic self-energy
which decreases with frequency  and gives rise to 
 SDW precursors at  $\Omega \geq v \xi^{-1}$
provided that $(g/g_0)^2 (T/v\xi^{-1})
\geq 1/4$ which is satisfied above $T^*$~\cite{joerg}.
Quantum SDW precursors can also appear at high frequencies 
provided that feedback effects
substantially reduce the spin damping~\cite{AC,Morr,KS}.
 In view of this nonuniversality, 
it is  possible that the broad maximum in $A(\omega)$ 
at high frequencies is actually 
a mixture of $d$-wave and SDW precursors.

To summarize, in this paper we discussed the thermal variation
of the fermionic spectral function
$A(\Omega)$ in the spin-fermion model for underdoped cuprates.
We found that in the strong coupling limit, the
fermionic Green's function
 near $(0,\pi)$ behaves  as 
$G^{-1} (\Omega) \propto  (\sqrt{\Omega} +\beta \sqrt{\omega_{sf}/2})$
 in the whole frequency range relevant to photoemission experiments. 
The frequency-independent term
is due to thermal spin fluctuations.
At low temperatures,  the $\sqrt{\Omega}$ term dominates and 
 yields the LEG in $A(\Omega)$ and  a broad maximum at larger
frequencies. As $T$ increases, spin fluctuations
progressively  acquire a  classical, quasistatic behavior.
This effect yields a rapid increase of $\beta$ with temperature
which in turn gives rise to a 
destruction of the LEG.
We found that the LEG disappears at 
$T =T^*\sim 0.6 \omega_{\rm sf} \xi^{1/\nu}$ 
 which is $150 -200K$
near $(0,\pi)$ and decreases as one moves towards zone diagonal.

It is our pleasure to thank J.C. Campuzano, L. Gorkov, 
M. Norman, M. Onellion, D. Pines, J. R. Schrieffer and  B. Stojkovi\'c
 for useful conversations. 
The research was supported by NSF DMR-9629839 (A.C)
and the STCS  through NSF-grant DMR91-20000
as well as the Deutsche Forschungsgemeinschaft (J.S.).
Part of this work has been done at the 
Aspen Center for Physics.

\end{document}